\documentclass[amsmath,amssymb,superscriptaddress,a4paper,aps,prl,reprint]{revtex4-1}

\usepackage{graphicx} % manages external pictures
\usepackage{lmodern} % enhanced version of Computer Modern fonts, available as Type 1
\usepackage{textcomp} %provides extra symbols like \textcelsius, etc.
\usepackage[T1]{fontenc} % utf8 encoding
\usepackage{gensymb} % \micro, \ohm, \degree, \celsius
\usepackage[seperr, load={}]{siunitx} % SI units

\def\be{\begin{equation}}
\def\ee{\end{equation}}

\newcommand*{\Pdet}{P_{\rm det}}
\newcommand*{\pow}{|s_{21}|^2}

\newcommand*{\Tb}{T_{\rm bath}}
\newcommand*{\IHpp}{I_H^{\rm pp}}

\newcommand*{\RT}{R_\mathrm{T}}

\newcommand*{\Gth}{G_{\rm th}}
\newcommand*{\Pin}{P_{\rm in}}

\newcommand*{\Gthnis}{G_{\rm th,NIS}}
\newcommand*{\Gthep}{G_{\rm th,ep}}

\newcommand*{\Qep}{\dot Q_{\rm ep}}
\newcommand*{\Qnis}{\dot Q_{\rm NIS}}
\newcommand*{\Pinj}{\dot Q_H}
\newcommand*{\Pzero}{\dot Q_0}
\newcommand*{\Pgen}{P_{\rm gen}}
\newcommand*{\Vb}{V_{\rm b}}

\begin{document}

\title{Fast electron thermometry towards ultra-sensitive calorimetric detection}
\author{S. Gasparinetti}
	\email{simone.gasparinetti@aalto.fi}
	\affiliation{Low Temperature Laboratory (OVLL), Aalto University, P.O. Box
	15100, FI-00076 Aalto, Finland}
\author{K. L. Viisanen}
	\email{klaara.viisanen@aalto.fi}
	\affiliation{Low Temperature Laboratory (OVLL), Aalto University, P.O. Box
	15100, FI-00076 Aalto, Finland}
	\author{O.-P.~Saira}
	\affiliation{Low Temperature Laboratory (OVLL), Aalto University, P.O. Box
	15100, FI-00076 Aalto, Finland}
	\author{T.~Faivre}
	\affiliation{Low Temperature Laboratory (OVLL), Aalto University, P.O. Box
	15100, FI-00076 Aalto, Finland}
	\author{M.~Arzeo}
	\affiliation{Quantum Device Physics Laboratory,
Department of Microtechnology and Nanoscience,
Chalmers University of Technology, SE-41296 G\"oteborg, Sweden}
	\author{M.~Meschke}
	\affiliation{Low Temperature Laboratory (OVLL), Aalto University, P.O. Box
	15100, FI-00076 Aalto, Finland}
	\author{J.~P.~Pekola}
	\affiliation{Low Temperature Laboratory (OVLL), Aalto University, P.O. Box
	15100, FI-00076 Aalto, Finland}

\date{\today}

\begin{abstract}
We demonstrate radiofrequency thermometry on a micrometer-sized metallic island
below 100 mK.
Our device is based on a normal metal-insulator-superconductor tunnel junction
coupled to a resonator with transmission readout.
In the first generation of the device, we achieve \SI{90}{\micro K/\sqrt{Hz}} noise-equivalent
temperature with \SI{10}{MHz} bandwidth.
We measure the thermal relaxation time of the electron gas in the island,
which we find to be of the order of \SI{100}{\micro s}.
Such a calorimetric detector, upon optimization, can be seamlessly integrated into superconducting circuits, with
immediate applications in quantum-thermodynamics experiments down to single quanta of energy.
\end{abstract}

\maketitle

\section{Introduction}
Thermometry is a key in studies of thermodynamics. When investigating large systems, it is often sufficient to monitor time-averaged temperatures, as the relative fluctuations are small. Then the bandwidth of the thermometer may not be an important figure of merit as such. In small systems, on the contrary, temporal statistical variations become increasingly important and it would be of great benefit to determine the effective temperature over time scales shorter than the relevant thermal relaxation time of the measured system. Despite the apparent lack of fast thermometers in mesoscopic structures, interesting experiments in thermal physics have been performed and are under way, including measurements of the quantum of heat conductance \cite{schwab00,meschke06,anthore13}, of Landauer's principle of minimum energy cost of erasure of a logic bit \cite{ciliberto12}, and of information-to-energy conversion in Maxwell's demons \cite{toyabe10,koski14}. Fast thermometry and calorimetry would tremendously expand the variety of phenomena to be explored, providing direct access to the temporal evolution of effective temperatures under non-equilibrium conditions, the energy-relaxation rates, and the fundamental fluctuations of the effective temperature in small systems. The observation of single quanta of microwave photons would eventually provide a way to investigate heat transport and its statistics in depth \cite{pekola13,gasparinetti14,silaev14}, for example in superconducting quantum circuits.

Here we demonstrate a significant step towards single-microwave-photon calorimetry beyond the seminal experiments in Refs.~\cite{nahum95,schmidt03,schmidt04,schmidt05}, down to electronic temperatures below \SI{100}{mK}.
%We measure the temperature of a micrometer-sized Cu island using a normal-metal-insulator (NIS) tunnel junction of \SI{22}{k\ohm} normal-state resistance and an rf-transmission readout. Our technique provides \SI{90}{\micro K/Hz^{1/2}} thermometry with a bandwidth of 10 MHz.
Our rf-transmission readout of a normal-insulator-superconductor (NIS) tunnel junction provides \SI{90}{\micro K/\sqrt{Hz}} thermometry with a bandwidth of 10 MHz.
%We characterize the thermal response of the island to controlled heat pulses and measure thermal relaxation times of the order of $\approx\SI{100}{\micro s}$.
Based on real-time characterization of the thermal response of the island, we conclude that the measured $\SI{100}{\micro s}$ relaxation time would allow us to detect a \SI{10}{mK} temperature spike in single-shot.
Our single-shot resolution has to be enhanced by
one order of magnitude in order to finally detect
a single \SI{1}{K} (\SI{20}{GHz}) photon impinging on an optimized
absorber.
%This spike is one order of magnitude higher than the one produced by a single \SI{1}{K} (\SI{20}{GHz}) photon impinging on an optimized absorber.
%The noise-equivalent power of our calorimeter is currently of the order of $10^{-18}$ \si{W/Hz^{1/2}}, one order of magnitude above the thermal-fluctuation limit.

%Our rf-transmission readout of a normal-insulator-superconductor (NIS) tunnel junction provides $\sim 100$ $\micro$K$/\sqrt{\rm Hz}$ thermometry with a bandwidth of 10 MHz. Based on real-time characterization of the thermal response of a micrometer-sized Cu island, we conclude that the measured $\approx\SI{100}{\micro s}$ relaxation time would allow us to detect single temperature spikes of 10 mK height. This is about one order of magnitude higher than the temperature rise expected for a typical single-photon absorption event in the microwave range.

\begin{figure}
\centering
	\includegraphics[width=\columnwidth]{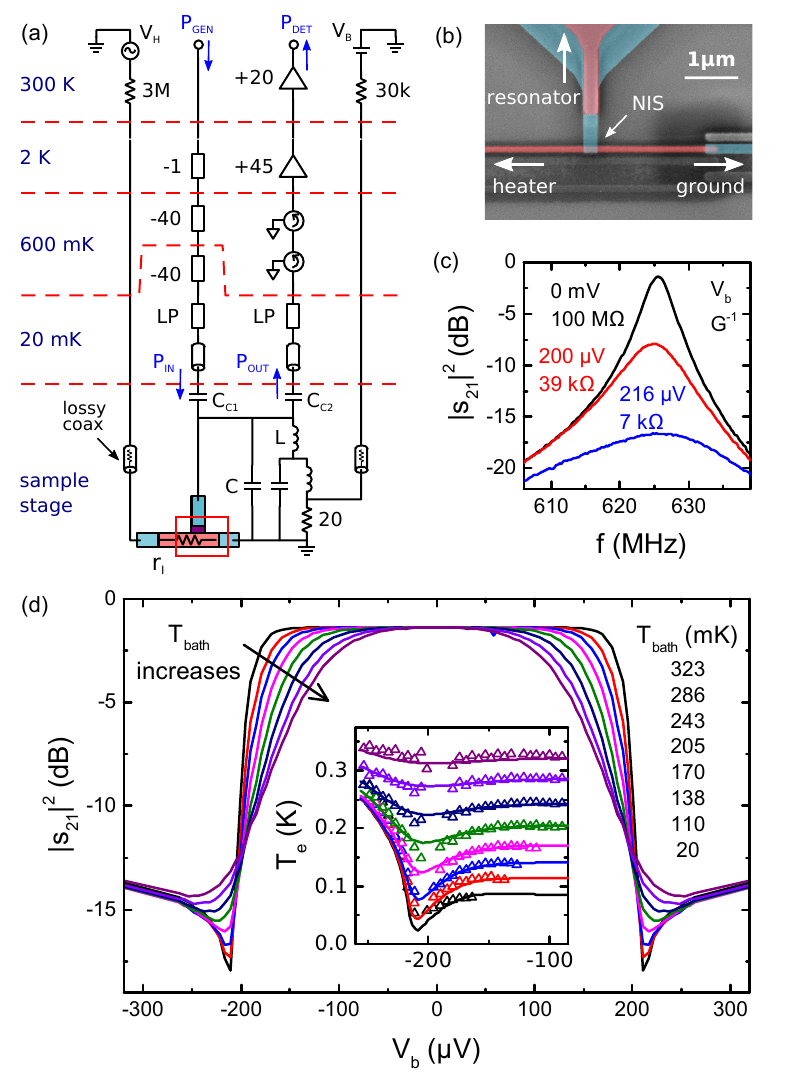}
	\caption{
	\textbf{The rf-NIS thermometer.}
(a) Schematic of the measurement circuit.
(b) False-color micrograph of a representative device (red: Cu, blue: Al), closing up on the NIS junction used as a thermometer.
(c) Small-signal transmittance $\pow$ versus frequency for three selected values of the voltage bias $V_b$; the corresponding differential resistance $G^{-1}$ of the NIS junction varies between \SI{7}{k\ohm} and \SI{100}{M\ohm}.
(d) Transmittance-voltage characteristics: $\pow$ versus $V_b$ for a set of bath temperatures $\Tb$ in the range of 20 to 323 mK. 
For each temperature, the transmittance at zero bias is taken as the \SI{0}{dB} reference.
Inset: Electronic temperature $T_e$ vs $V_b$ for different values of $\Tb$.
The experimental points (triangles) are obtained from the data of the main panel using Eqs.~\eqref{eq:s21} and \eqref{eq:G}.
The predictions of a thermal model taking into account electron-phonon and tunneling heat conductance \cite{suppl} are shown for comparison (solid lines).
}

\label{fig:intro}
\end{figure}

\section{Characterization}
Our technique relies on the temperature-dependent conductance of the NIS junction \cite{rowell76, nahum93, giazotto06}.
In the standard dc configuration, the high impedance of the junction, together
with stray capacitance from the measurement cables, limits its bandwidth to the
kHz range. In order to enable a fast readout, we embed the NIS junction in an $LC$ resonant circuit \cite{schmidt03}. Similar techniques are routinely used for the fast readout of high-impedance nanodevices, including single-electron transistors \cite{schoelkopf98} and quantum point contacts \cite{qin06,reilly07}.
%***Low-impedance NIS junctions have also used for bolometry in previous experiments \cite{nahum95,schmidt05}, with advantages in terms of both sensitivity and bandwidth.

Our sample consists of a 25 nm thick, 100 nm wide and 20 $\micro$m long Cu island connected to Al
leads via two clean normal metal-superconductor (NS) contacts and a NIS junction with normal-state resistance $R_T = \SI{22}{k\ohm}$.
A schematic of our measurement set-up is shown in Fig.~\ref{fig:intro}(a) and a close-up, false-color micrograph of the device is shown in Fig.~\ref{fig:intro}(b).
The device is fabricated on top of an oxidized silicon substrate by standard
electron-beam lithography, three-angle metal evaporation with in-situ Al oxidation, and
liftoff.
The NIS probe is embedded in an $LC$ resonator formed by a $L=\SI{80}{nH}$ surface-mount inductor, which together with the stray capacitance $C=\SI{0.5}{pF}$ and coupling capacitors $C_{\mathrm{C1}}=0.1~\mathrm{pF}$, $C_{\mathrm{C2}}=0.2~\mathrm{pF}$ gives a resonant frequency $f_0=\SI{625}{MHz}$.
A bias tee allows a dc voltage bias $V_b$ to be applied to the NIS
junction without interfering with the resonator readout.
Of the two NS contacts, one is grounded at the sample stage,
while the other is used to feed a heating current to the island.
The total resistance between the normal electrode of the NIS junction and the ground, including the resistance of the NS contact, was measured to be \SI{360}{\ohm}.
%The total resistance of the island is $r_I=\SI{360}{\ohm}$, of which less than $10\%$ lies between the NIS probe and the grounding NS contact.

We probe the resonator, coupled to input and output ports via the capacitors $C_{C1}$ and $C_{C2}$, by measuring the transmittance $|S_{21}|^2=P_{\rm det}/P_{\rm gen}$, see Fig.~1(a).
For the time-resolved measurements described in the following, the signal is demodulated at the carrier frequency and recorded with a fast digitizer.
The rf input line is attenuated by 80 dB below 2 K
before reaching the sample stage. Two circulators in series ensure at least \SI{45}{dB} isolation between the
resonator output and a low-noise high-electron-mobility-transistor (HEMT) amplifier mounted on the 2 K plate.
The bias and heating lines are filtered by a 2 m long lossy coaxial line (Thermocoax).
Sample and resonator are enclosed in an rf-tight, indium-sealed \cite{saira12} copper box mounted at
the base plate of a dilution refrigerator cooled down to 20 mK.
The base plate temperature $\Tb$ is measured by a calibrated RuOx thermometer.
%providing strong attenuation at frequencies above 1 GHz.
%Furthermore, each line has around \SI{200}{\ohm} dc resistance and \SI{1}{nF} distributed capacitance, thereby realizing a distributed $RC$ filter whose time constant is \SI{0.2}{\micro s}.

At low input power, the resonator probes the differential conductance $G= \partial I / \partial V_b$ of the junction at the bias point $V_b$.
Figure 1(c) shows how the resonance peak responds to changes in $V_b$.
The transmittance of the resonator at resonance
is given by \be
|s_{21}| = 2 \kappa \frac{G_0}{G+G_0} \ , \label{eq:s21}
\ee
with $\kappa = C_\mathrm{C1}C_\mathrm{C2}/(C_\mathrm{C1}^2+C_\mathrm{C2}^2)$ and $G_0 = 4 \pi^2 (C_{C1}^2+C_\mathrm{C2}^2) Z_0 f_0^2$ (here $Z_0=\SI{50}{\ohm}$ is the transmission line impedance and $f_0$ is the resonance frequency).
By measuring $\pow$ at $V_b=0$ and $V_b \gg \Delta/e$, where $G \ll G_0$ and $G \approx R_T^{-1}$, respectively, we estimate $G_0 \approx \SI{22}{\micro S}$.
For each curve in Fig.~1(c) we note the corresponding differential resistance $G^{-1}$, emphasizing the high sensitivity of the readout at impedances of the order of $1/G_0 \approx \SI{50}{k\ohm}$. At that impedance the bandwidth, defined as the FWHM of the resonance curve, is \SI{10}{MHz} and the loaded $Q$ factor is $62.5$.
In the following we will probe the resonator at resonance.
% Shifts in $f_0$ upon changing the bath temperature $\Tb$ were found to be negligible in the temperature range of interest. 

% \textbf{Continuous-wave characterization} --
With the calibrated resonator parameters $\kappa$ and $G_{\mathrm{0}}$, 
a measurement of the transmitted power provides
the same information as the conventional current-voltage characteristics of an NIS junction.
In particular, such a measurement makes it possible to infer the electronic temperature $T_e$ in the Cu island. To extract $T_e$ from $\pow$, we first convert $\pow$ into $G$ using
%the resonator response function
\eqref{eq:s21} and then compare the result to the expression for the conductance of the NIS junction
\be
G =  \frac1{R_T k_B T_e} \int dE N_S(E) f(E-eV_b) \left[ 1 -
f(E-eV_b)\right] \ , \label{eq:G}
\ee
where $k_B$ is the Boltzmann constant, $e$ the electron charge, $N_S(E)=\left|\Re{\rm e} \left(E/\sqrt{E^2-\Delta^2}\right)\right|$ the normalized Bardeen-Cooper-Schrieffer
superconducting density of states, $f(E)=\left[1+\exp(E/k_BT_e)\right]^{-1}$
the Fermi function, and $\Delta$ is the superconducting gap.
Notice that the temperature of the superconducting electrode does not appear in \eqref{eq:G}; this is a well-known property of the NIS thermometer \cite{pothier97}. Moreover, at the low bias voltages of the thermometer, the backflow of heat from the superconductor is not significant at these temperatures \cite{muhonen12}.

In Fig.~1(d) we plot $\pow$ as a function of $V_b$ for a set of bath temperatures $\Tb$ in the range of 20 to 325 mK.
The corresponding $T_e$ versus $V_b$, as extracted from the traces in the main panel, is plotted in Fig.~1(d), Inset (triangles).
We have excluded points around $V_b=\Delta/e$ where the first-order temperature sensitivity vanishes.
At base temperature $\Tb = \SI{20}{mK}$ we find that $T_e \approx \SI{85}{mK}$. This saturated $T_e$ corresponds to a spurious injected power $\dot Q_0 \approx \SI{400}{aW}$ \cite{suppl},
which we ascribe to imperfect shielding of blackbody radiation as well as low-frequency noise in the dc lines and in the ground potential.
The dependence of $T_e$ on $V_b$, most pronounced for the
lowest-temperature traces, is due to heat transport across the NIS junction. In
particular, cooling is expected to take place when $V_b
\approx \Delta/e$ \cite{nahum94}, and heating when $V_b \geq \Delta/e$.
Conversely, at high temperatures, $T_e$ closely follows $\Tb$, as the electron-phonon heat conductance provides a strong thermal anchoring to the electrons in the Cu island. The agreement between $T_e$ and $\Tb$ establishes the validity of the rf-NIS electron thermometry.
Furthermore, our data are quantitatively accounted for by a simple thermal model which takes the most relevant heat flows into account \cite{suppl}.
The calculated $T_e$ (solid lines) agrees well with the measured ones, except in the vicinity of the optimal cooling point, where only a modest cooling is observed if compared to the theoretical prediction. This behavior can be ascribed to local overheating of the superconductor \cite{rajauria09}, not included in the model. %, and is typically observed unless particular care is taken to evacuate excess quasiparticles at the junction.

\begin{figure}
\centering
\includegraphics[width=\columnwidth]{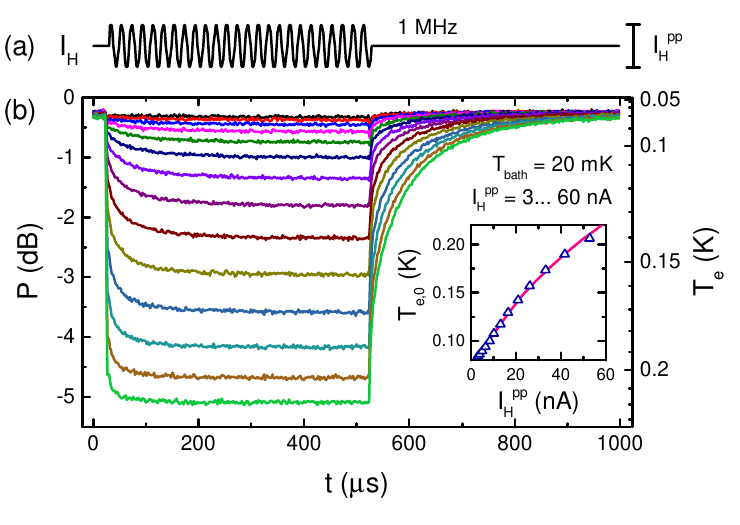}
	\caption{\textbf{Time-resolved thermometry.}
(a) Amplitude-modulated sinusoid used to drive the heating pulse (the frequency is not to scale) and (b) real-time response of the thermometer, obtained by recording the transmitted power $P$ versus time for different values of the heating-pulse amplitude $\IHpp$.
The conversion from $P$ into absolute electronic temperature $T_e$ is displayed on the right axis.
Inset: $T_e$ at the end of the heating pulse ($t=\SI{520}{\micro s}$) versus $\IHpp$ (triangles). The prediction of the thermal model \cite{suppl} is shown for comparison (solid line). All the traces are taken at base temperature by averaging over $10^4$ heating cycles and the voltage bias is $V_b=\SI{0.17}{mV}$.
%(c) Sketch of the thermal model: electrons in the Cu island, whose heat capacity is $\mathcal C$, are heated by the Joule power $\dot Q_H$, which adds up to the spurious $\dot Q_0$. Heat is carried away by phonons ($\dot Q_{\rm ep}$) as well as electrons tunneling through the NIS junction ($\dot Q_{\rm NIS}$). Balancing the heat flows yields the steady-state $T_e$.
}
	\label{fig:traces}
\end{figure}

%\textbf{Time-resolved measurements} --
\section{Time-resolved measurements}
We demonstrate the real-time capability of our thermometer by measuring the thermal relaxation of the electron gas in the Cu island in response to a Joule heating pulse.
The heating pulse is generated by feeding an amplitude-modulated sinusoid of frequency $f_H = \SI{1}{MHz}$ to a large bias resistor, resulting in an ac heating current of peak-to-peak amplitude $\IHpp$. As $f_H$ is much faster than the measured thermal relaxation rates (see the following), the island reacts to a time-averaged heating power $\dot Q_H \propto (\IHpp)^2$ when the heating is on.
The time-domain response of the thermometer to the heating pulse is shown in Fig.~2(b) at base temperature, for a fixed $V_b$ and different values of $\IHpp$. The left axis indicates the instantaneous power recorded by the digitizer. This power is converted into temperature using a similar procedure as in Fig.~1(d), Inset, and the corresponding scale is noted on the right axis. The temperature reached by the island at the end of the heating pulse is plotted in Fig.~2(b), Inset as a function of $\IHpp$ (triangles), in good agreement with the prediction of the thermal model (solid line).
% The latter takes into account heat relaxation via electron-phonon coupling ($\dot Q_{\rm ep}$) and via electron tunneling through the NIS junction ($\dot Q_{\rm NIS}$) \cite{suppl}.
From Fig.~2, we see that the thermal response of the island is not instantaneous; instead,
a finite-time relaxation is observed after the rising and falling edge of the pulse. %, corresponding to different steady-state temperatures.
%A quantitative characterization of the thermal relaxation is the subject of the following analysis.

With constant heat input and when $T_e$ is not far from its steady-state value $T_{e,0}$, the heat equation governing the temperature deviation $\delta T = T_e-T_{e,0}$ can be written as
\be \label{e1}
\mathcal C\frac{d\delta T}{dt} = - \Gth\delta T,
\ee
where $\mathcal C$ is the electronic heat capacity of the island and $\Gth$ the thermal conductance to its heat bath.
Equation \eqref{e1} tells that $T_e$ relaxes to $T_{e,0}$ exponentially with the relaxation time $\tau = \mathcal C/\Gth$, where $\mathcal C$ and $\Gth$ are to be evaluated at $T_e=T_{e,0}$. Even after a large change in the heating power [beyond the linear-response regime described by \eqref{e1}], the final approach to the new $T_{e,0}$ obeys this exponential law.
% The same Eq.~\eqref{e1}, evaluated at the proper $T_{e,0},$ describes relaxation to a hotter temperature after the pulse begins, just as well as relaxation to a colder temperature after the pulse ends.
The value of $\mathcal C$ is ideally given by the standard expression for a Fermi electron gas, $\mathcal C=\gamma \mathcal V T_{e,0}$, where $\gamma = 71$ JK$^{-2}$m$^{-3}$ \cite{roberts78} and $\mathcal V$ is the volume of the island (in our case, $\mathcal V=\SI{0.05}{\micro m^3}$). On the other hand, $\Gth$ is determined by the sum of
all relevant parallel heat conductances. In the present case we expect the electron-phonon heat conductance $\Gthep$ and the tunneling heat conductance through the biased NIS junction $\Gthnis$ to be the dominant contributions.
Thermal conductivity through the clean NS contacts can be neglected \cite{peltonen10} and
%, for the following reasons: (a) Andreev reflection blocks the heat flow within the gap, (b) residual heat conduction due to quasiparticle excitations is effectively frozen out at low temperatures, and (c) our superconducting lines are much longer than the coherence length in order to suppress the inverse proximity effect 
photonic heat conductance is also negligible for our sample at these temperatures, due to the mismatch of the relevant impedances \cite{timofeev09}.
Measurements of the heat conductance out of a metallic island were recently reported in \cite{govenius14}.
The standard expression for $\Gthep$ is quoted as $\Gthep =  5\Sigma \mathcal V T_e^4$ \cite{wellstood94}; however, other power laws in $T_e$ have also been reported for experiments on Cu islands \cite{taskinen04,karvonen05}.
The tunneling heat conductance is given by
$
\Gthnis = - \frac{1}{e^2 R_T k_B T^2} \int_{-\infty}^\infty dE N_S(E) (E- eV)^2 f(E-eV)[1-f(E-eV)] \ .
$
For our relatively large island and according to these expressions, we expect $\Gthep \gg \Gthnis$ when the junction is biased far from the gap and $\Gthep \approx \Gthnis$ when $V_b$ approaches $\Delta/e$. However, as indicated by the data in Fig.~1(d), Inset, the cooling performance of the NIS junction is degraded when $V_b \approx \Delta/e$, possibly implying a weaker $\Gthnis$ than predicted by the model.
Finally, it should be mentioned that the electron-phonon relaxation times reported in \cite{schmidt04,taskinen04} were longer than those expected based on the expressions above. In addition to a non-ideal $\Gthep$, this may suggest a one order of magnitude larger heat capacity than described by the Fermi gas model, possibly due to magnetic impurities in the metal film \cite{pobell,anthore03}. Furthermore, overheating of the local phonon bath, considered in a recent experiment \cite{pascal13}, may also lead to longer relaxation times, due to the additional series thermal resistance between the local phonon bath and the thermalized substrate phonons.

\begin{figure}
\centering
	\includegraphics[width=\columnwidth]{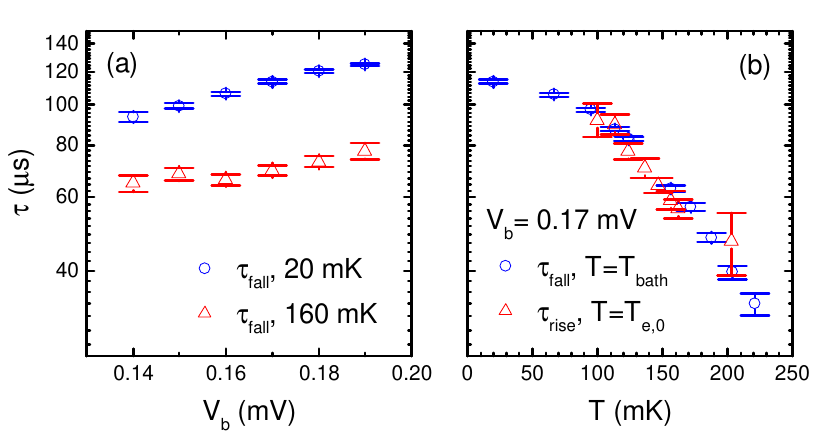}
	\caption{\textbf{Thermal relaxation times.}	
%(a--d) Thermal relaxation traces (circles, squares, triangles). The traces are shifted by their baseline after relaxation, scaled and plotted on a logarithmic scale. They are also horizontally offset by \SI{150}{\micro s} for clarity. The full lines are exponential fits of the form $A\exp(t/\tau)+B$ to the data. See also \cite{suppl}. The data in panels (a,b) correspond to the rising (a) and falling
%edges (b) of selected traces in Fig.~2.
%Panels (c,d) present similar traces obtained for different values of the voltage bias $V_b$ (c), and at different bath temperatures $\Tb$ (d).
(a) Thermal relaxation time $\tau$ versus voltage bias $V_b$ for two different values of the bath temperature $\Tb$.
(b) Temperature dependence of $\tau$, as estimated from relaxation after the falling edge (circles, the $x$ axis is $\Tb$) as well as the rising edge of the pulse (triangles, the $x$ axis is the temperature $T_{e,0}$ at the end of the pulse). The error bars are obtained from the fits (see \cite{suppl}).
}
	\label{fig:taus}
\end{figure}

We estimate the thermal relaxation times $\tau_{\rm rise}$ and $\tau_{\rm fall}$ by fitting an exponential function to the tails of the relaxation traces observed in Fig.~2 after the rising ($\tau_{\rm rise}$) and falling edge ($\tau_{\rm fall}$) of the heating pulse. More details on the fitting procedure are given in \cite{suppl}. As we increase the pulse amplitude $\IHpp$, we observe a decrease in $\tau_{\rm rise}$, which is consistent with thermal relaxation to a higher temperature. On the other hand, $\tau_{\rm fall}$ does not depend on $\IHpp$, as expected due to the fact that the relaxation temperature stays the same. We have repeated the measurements of Fig.~2 while varying the bias voltage $V_b$ and the bath temperature $\Tb$. The corresponding relaxation times $\tau$ are shown in Fig.~\ref{fig:taus}.
%The tails are obtained from the raw data by subtracting the steady-state-temperature baseline from each trace. They have been normalized, horizontally offset for clarity, and plotted in a semilogarithmic scale in order to highlight the exponential decay. The full lines are fits of an exponential function to the tails \cite{long_tau_note}.
%The tails in panels (a,b) refer to relaxation after the rising (a) and falling edge (b) of heating pulses of different amplitude $\VHpp$. As $\VHpp$ is increased, relaxation after the rising edge gets faster as $T_{e,0}$ increases; on the other hand, no change is observed in the tails after the falling edge, as $T_{e,0}$ stays the same.
%In panel (c), we vary the bath temperature $\Tb$ and see that the relaxation gets faster as $\Tb$ is increased.
%In panel (d), we vary the bias voltage $V_b$. The observed time constant stays approximately the same, regardless of the fact that $G$ changes by over two orders of magnitude across the given $V_b$ range.
%Finally, in panels (e,f), we plot the values of $\tau$ as obtained from the fits.
In panel (a) we show the dependence on $V_b$ for two different values of $\Tb$. The measured $\tau$ at base temperature is of the order of \SI{100}{\micro s} and it increases by some 20\% as $V_b$ approaches $\Delta/e$. This increase may well be due to a decrease in $\Gthep$ upon cooling of the island [compare Fig.~1(d), Inset].
In panel (b) we show the temperature dependence of $\tau$, obtained in two independent ways. We first measured $\tau_{\rm fall}$ while varying $\Tb$ (circles) and then $\tau_{\rm rise}$ while varying $\IHpp$ (triangles). $\tau_{\rm fall}$ is plotted against $\Tb$ and $\tau_{\rm rise}$ is plotted against $T_{e,0}$ at the end of the pulse, estimated as in Fig.~2(b). The agreement between the two series is remarkable. The saturation of $\tau$ at low $\Tb$ is also consistent with the saturated $T_e$ observed in Fig.~1(d), Inset. From the measured $\tau$ we estimate a heat capacity $\mathcal C = 2\cdot 10^{5}k_{\mathrm{B}}=\SI{3}{aJ/K}$, one order of magnitude larger than the expected value for a Cu island of the size and temperature in this experiment.
At higher temperatures $\tau$ is predicted to scale as $T_{e,0}^{-3}$ provided $\Gth \approx \Gthep$ and both $\mathcal C$ and $\Gthep$ follow the theory predictions. The data presented here are not conclusive in this respect, due to the saturation of $T_{e,0}$ at low $\Tb$ and to the narrow temperature range considered.
This range is not limited by the bandwidth of our thermometer, but rather by a transient that we observe after terminating the heat pulse, possibly due to the heavy low-pass filtering applied to the heating line. For this reason, we refrain from presenting data points with $\tau \lesssim \SI{20}{\micro s}$ and leave the study of relaxation times down to $\SI{1}{\micro s}$ and below to future investigation.

% This range is not limited by the bandwidth of our thermometer, but rather by the low-pass filtering applied to the heating line, which makes it difficult to sharply terminate the heating pulse. Due to this limitation, we refrain from fitting traces with $\tau < \SI{20}{\micro s}$ and leave the study of relaxation times down to $\SI{1}{\micro s}$ and below to future investigation.

\section{Noise and responsivity}
%in the frequency range of 1 Hz to 1 MHz, at different voltage biases, bath temperatures and input powers, and using different noise-equivalent bandwidths when digitizing the signal
%The nominal NT of out HEMT amplifier, specified at 20 K and 550 MHz, is 9.3 %K.
We have performed an extensive characterization of the responsivity and noise of the thermometer readout. Our first set of experiments, presented above, were performed at low input powers corresponding to a voltage modulation amplitude across the NIS junction of the order of $\SI{1}{\micro V}$. In this case, the readout probes the local differential conductance of the junction. Accordingly, the theoretical responsivity $\mathcal R = \partial P_{\rm det}/ \partial T_e$ of the thermometer is $\mathcal R \propto P_{\rm gen}  (\partial \pow / \partial G) (\partial G / \partial T_e)$.
We evaluate the noise-equivalent temperature (NET) as $ ( \delta P_{\mathrm{det}}/\delta T )^{-1}(\sqrt{S_{P_\mathrm{det}P_\mathrm{det}}})$, where $S_{\mathrm{P_\mathrm{det}P_\mathrm{det}}}$ is
the measured noise spectral density of the detected power $\Pdet$. At an electron temperature of 80 mK and at the optimal bias point of 0.17~mV, we obtain our best NET of \SI{90}{\micro K/\sqrt{Hz}}. We always find an essentially white noise spectrum, with a corner frequency for $1/f$ noise of the order of a few Hz.
%Due to additional reflection at the output port of our
%sample box, the measured NET is larger than the optimal achievable
%value and the effective noise temperature of the chain is 200~$\mathrm{K}$,
%despite the noise temperature of our amplifier being close to its nominal value of \SI{9.3}{K}. 
%We predict that a five-fold improvement in NET can be achieved by using a perfectly matched sample box.

The thermometer readout was amplifier-limited. We characterize the noise of the rf readout chain by the system noise temperature $T_\mathrm{sys}$ referred to the output port of the sample box. In this case (see supplement for details~\cite{suppl}), one has $S_{P_\mathrm{det}P_\mathrm{det}} \approx 4 G k_B T_\mathrm{sys} P_\mathrm{det}$, where $G = P_\mathrm{det}/P_\mathrm{out}$ is the total gain of the amplification chain. Using power-dependent features of the NIS-junction-loaded resonator as markers, we estimate $G = 55 \pm 1$~dB and $T_\mathrm{sys}$ = $62 \pm 15$~K. The discrepancy between $T_\mathrm{sys}$ and the nominal noise temperature of the HEMT amplifier, 13.3 K at 640~MHz, suggests an insertion loss of the order of 7 dB between the resonator and the amplifier.

Assuming the heat conductance $\Gth$ to be dominated by electron-phonon interaction [as indicated by the steady-state measurements of Fig.~1(d)], the noise-equivalent power (NEP) is given by $\mathrm{NEP} = \mathrm{NET} \, G_{\rm th} = \SI{2.5e-18}{W/\sqrt{Hz}}$. This figure is one order of magnitude above the
thermal fluctuation noise limit
$\mathrm{NEP}_{\rm th}=\sqrt{4 k_B T_e^2 \Gth}= \SI{1e-19}{W/\sqrt{Hz}}$.

\begin{figure}
\centering
	\includegraphics[width=\columnwidth]{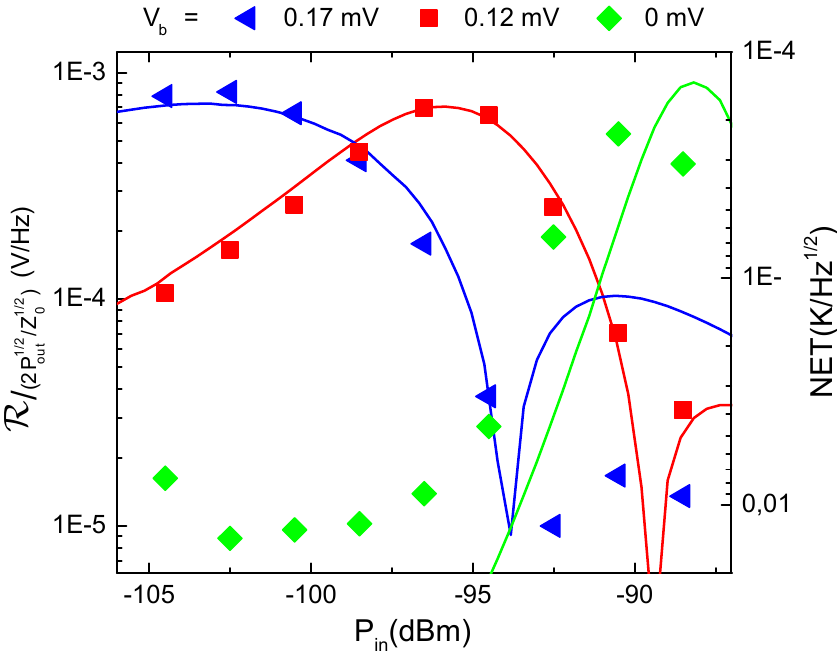}
	\caption{\textbf{Power optimization.} Normalized responsivity $\mathcal{R}$ (left axis) and corresponding noise-equivalent temperature (right axis) versus $P_{\rm in}$ for three selected bias voltages (symbols), measured at \SI{150}{mK}. Numerical simulations are shown for comparison (solid lines, see \cite{suppl} for details). }
	\label{fig:sens}
\end{figure}

One may ask whether the NET figure given above can be significantly improved by operating
the rf-NIS thermometer at higher
input powers, \emph{i.e.}, beyond the linear regime.
In Fig.~\ref{fig:sens} we compare the responsivity and NET of our thermometer at different bias voltages and as a function of the power fed to the input line. The data (symbols) were taken in a separate cooldown using an equivalent setup and a sample with $R_T=\SI{28}{k\ohm}$. The optimal power increases as the bias point is shifted towards zero bias. Importantly, a sensitivity close to the global optimum (\SI{144}{\micro K/\sqrt{Hz}} for this sample at $\Tb = 150$~mK) is reached over a broad range of bias voltages by a suitable choice of probing power. This feature can be understood by considering the combined contribution of the dc bias and the rf drive to the instantaneous voltage across the junction, and the fact that the responsivity of the NIS thermometer is concentrated in a narrow voltage range slightly below the superconducting gap edge. Indeed, full numerical simulations (solid lines) confirm this behavior. 

\section{Outlook}

In summary, we have demonstrated an electronic thermometer with promise for ultralow-energy calorimetry, operating below \SI{100}{mK}, with \SI{90}{\micro K/\sqrt{Hz}} noise-equivalent temperature and 10 MHz bandwidth. We have measured thermal relaxation times up to 100 $\micro$s, in line with 1.6 -- \SI{20}{\micro s} measured by other methods at higher temperatures \cite{schmidt04,taskinen04}. These figures already enable single-shot detection of an energy-absorption event producing a \SI{10}{mK} temperature spike.
Such a spike could be generated, for instance, by a single THz photon impinging of an absorber of reduced volume, as well as by a multi-photon wave packet in the C and X band used for superconducting-quantum-bit readout \cite{wallraff2004,houck2007,bozyigit2010}. In absolute terms, the NEP performance of our device still lags behind that of state-of-the-art transition-edge sensors \cite{karasik11,karasik11a} and semiconductor bolometers \cite{komiyama00,komiyama11}, which routinely achieve NEPs of the order of $10^{-20}$ \si{W/\sqrt{Hz}}. However, most of these devices are intended for detection of THz radiation, while our primary focus is on microwave photons. In the microwave domain, our approach presents some advantages; in particular, our sensor can be straightforwardly integrated in superconducting coplanar waveguides, acting as a lumped-element resistor whose impedance can be made to be of the order of \SI{50}{\ohm}.

Our current device and set-up leave room for improvement. Calculations indicate that the sensitivity of a fully optimized NIS thermometer can reach ${\rm NET_{\mathrm{opt}}} = \sqrt{2.72 \mathrm{e}^2 T_\mathrm{sys} \RT/k_\mathrm{B}}$. Using the parameters for our primary sample ($\RT = $ \SI{22}{k\Omega}) and present set-up ($T_\mathrm{sys} = 62$~K), this formula yields ${\rm NET_{\mathrm{opt}}} = \SI{83}{\micro K/\sqrt{Hz}}$, to be compared with our experimental value of \SI{90}{\micro K/\sqrt{Hz}}. We conclude that the impedance matching between the NIS junction and the transmission line realized by the resonator was close to optimal. Instead, the system noise temperature could be lowered by more than an order of magnitude by reducing losses between the sample box and the amplifier and by employing an amplifier with a lower noise temperature as the first stage; a Josephson parametric amplifier \cite{castellanos-beltran07} is one such choice.
The energy resolution of our detector can be estimated as $\delta E =  \mathcal C \, \delta T= {\rm NET} \, \mathcal C \, \tau^{-1/2}$. For the present case this gives $\delta E =2.3\cdot 10^{-20}~\rm J$, corresponding to a photon of frequency $\delta E/h = \SI{34}{THz}$. 
The measured sample was not optimized for obtaining a small
energy resolution; instead, we aimed at a strong coupling
between the island and the phonon bath.
In order to boost energy resolution, the size of the island can be
made significantly smaller, which is the next step toward improving this device.
When the noise is limited by thermal fluctuations, we can write $\delta E = \sqrt{4k_\mathrm{B}\mathcal V\gamma ^2/(5\Sigma\tau)}$.
For a sample with 50 times smaller island
limited by thermal fluctuations,
the energy resolution
at 80~$\mathrm{mK}$, assuming 100~$\mathrm{\mu s}$ relaxation time,
is $\delta E/h = \SI{30}{GHz}$.
Since $\tau$ increases strongly with decreasing temperature,
lowering the island temperature is another key point.
Optimized as indicated, our detector will facilitate a series of experiments of fundamental relevance in classical and quantum thermodynamics, as well as calorimetric measurements of dissipation down to single microwave photons in superconducting quantum circuits.

%The optimal energy resolution is expected when the heat conductance is dominated by electron tunneling through the NIS junction, rather than by electron-phonon interaction, as in the present case \cite{nahum95}.
%With such a small volume, electronic heat conduction through the NIS junction may become significant even at voltages below the gap, calling for a careful tradoff between high sensitivity and low heat conductance.
%we envisage NEPs in the range of $10^{-19}$ \si{W/\sqrt{Hz}}, limited by thermal fluctuations, to be attainable with the technology presented here. 

\section{Acknowledgements}

We would like to thank A.~Adamyan, S.~Kubatkin, J.~Govenius, R.~Lake and J.~Peltonen for useful discussions and S.~Kafanov for technical assistance at an early stage of the project. This work has been supported in part by the Academy
of Finland (project no.~139172) and its LTQ
(project no.~250280), and the
European Union Seventh Framework Programme INFERNOS (FP7/2007-2013) under grant agreement no.~308850.
S.~G.~acknowledges financial support from the Finnish National Graduate School in Nanoscience (NGS-NANO) and from the Aalto Doctoral
Programme in Science.

\clearpage
\onecolumngrid

\begin{center}
\large \textbf{Supplemental Material for ``Fast electron thermometry towards ultra-sensitive calorimetric detection''}
\end{center}

\section{Thermal model}

In order to estimate the steady-state electronic temperature $T_e$, we numerically solve a power-balance equation of the conventional form
\be
\Qep(T_e,\Tb) + \Qnis(V_b,T_e) + \Pinj(V_H) + \dot Q_0 = 0 \ . \label{eq:Qdots}
\ee
Here, we take temperature relaxation via electron-phonon coupling to be given by the standard expression $\Qep = \Sigma \mathcal V (T_e^5 - \Tb^5)$, where
$\Sigma=\SI{2e9}{W m^{-3} K^{-5}}$ is the electron-phonon interaction constant, $\mathcal V$ is the island volume and we assume the local phonons to be termalized at the bath temperature $\Tb$. The heat flow into the island due to electron tunneling through the NIS junction is given by
\be
\Qnis = - \frac{1}{e^2 R_T} \int_\Delta^\infty dE N_S(E) \left[ (E- eV_b)f_N(E-eV_b) + (E+eV_b )f_N(E+eV_b) - 2 E f_S(E)\right] \ , \label{eq:Qnis}
\ee
where $V_b$ is the voltage bias, $R_T=\SI{22}{k\ohm}$ is the tunneling resistance of the junction, $f$ is the Fermi function, the subscripts N and S refer to the normal and superconducting electrode, respectively, and $N_S$ is the BCS density of states. The last two terms in \eqref{eq:Qnis} can be neglected provided $k_BT_{N,S}<0.3\Delta$, where $\Delta$ is the zero-temperature superconducting gap.
The power fed through the heating line is $\Pinj(V_H)=V_H^2 r_I/R_H^2$, where $V_H$ is the heating voltage, $R_H=\SI{3}{M\ohm}$ is the room-temperature bias resistor and $r_I=\SI{360}{\ohm}$ the total resistance of the island. Finally, we assume that some spurious, constant heating power $\Pzero$ is delivered to the island due to imperfect filtering.
There are two free parameters in the model: $\Delta$ and $\Pzero$.
In particular, the value $\Delta = \SI{213}{\micro eV}$, in good agreement with other measurements on thin Al films, can be inferred from the crossing point of the curves in Fig.~1(d) in the main text. The value $\Pzero = \SI{400}{aW}$ essentially determines the value of $T_e$ observed at low $\Tb$. All the theoretical curves in Fig.~1(d), Inset in the main text were produced using these values for $\Delta$ and $\Pzero$. 

\section{Analysis of thermal relaxation times}

\begin{figure}
\centering
	\includegraphics{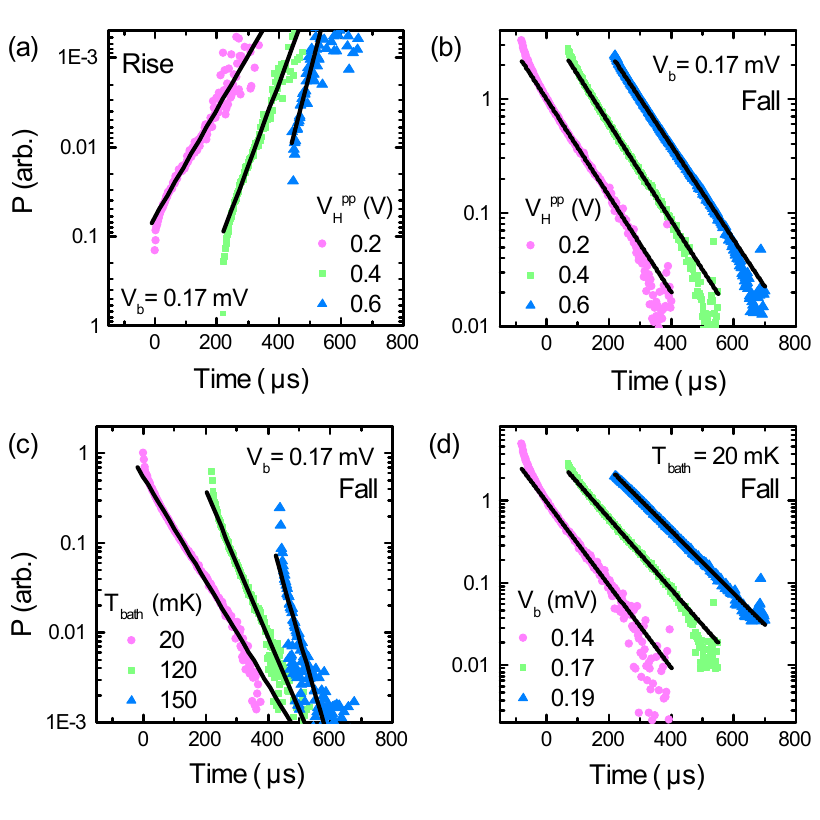}
	\caption{Thermal relaxation traces (circles, squares, triangles). The traces are shifted by their baseline after relaxation, scaled and plotted on a logarithmic scale. They are also horizontally offset by \SI{150}{\micro s} for clarity. The full lines are exponential fits of the form $A\exp(t/\tau)+B$ to the data. The data in panels (a,b) correspond to the rising (a) and falling edges (b) of selected traces in Fig.~2 of the main text.
Panels (c,d) present similar traces obtained at different bath temperatures $\Tb$ (c) and for different values of the voltage bias $V_b$ (d). All the traces are obtained by averaging over $2\cdot 10^5$ heating cycles.
}
\label{fig:tails}
\end{figure}

In Fig.~\ref{fig:tails}, we present relaxation tails obtained from measurements similar to those presented in Fig.~2 in the main text.
The tails are obtained from the raw data by subtracting the steady-state-temperature baseline from each trace. They have been normalized, horizontally offset for clarity, and plotted in a semilogarithmic scale in order to highlight the exponential decay. The full lines are fits of an exponential function to the tails.
The tails in panels (a,b) refer to relaxation after the rising (a) and falling edge (b) of heating pulses of different amplitude $\IHpp$. As $\IHpp$ is increased, relaxation after the rising edge gets faster as $T_{e,0}$ increases; on the other hand, no change is observed in the tails after the falling edge, as $T_{e,0}$ stays the same.
In panel (c), we vary the bath temperature $\Tb$ and see that the relaxation gets faster as $\Tb$ is increased.
In panel (d), we vary the bias voltage $V_b$. The observed time constant stays approximately the same, regardless of the fact that $G$ changes by over two orders of magnitude across the given $V_b$ range.

\section{Long time scale in the relaxation traces}

\begin{figure}[b]
\centering
\includegraphics{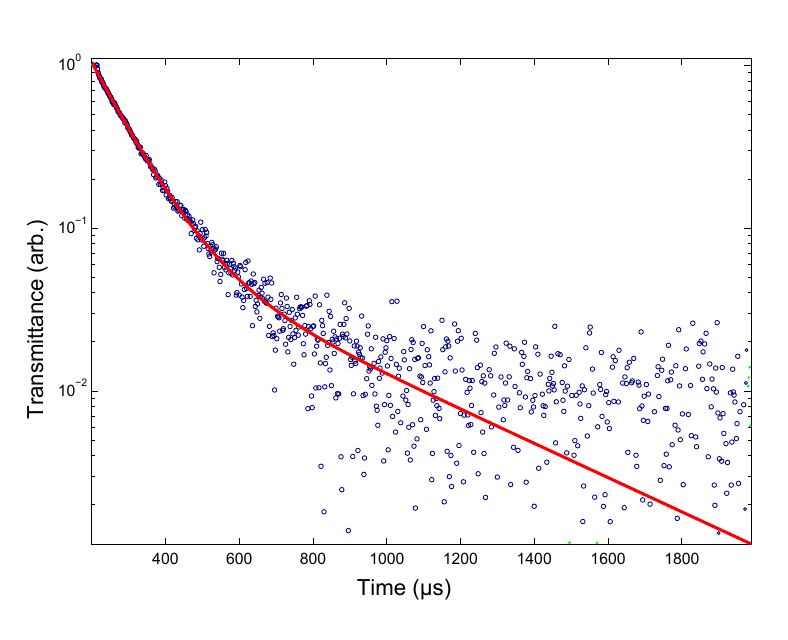}
\caption{Detail of a \SI{10}{ms} long time trace taken under the same conditions as in Fig.~3 of the main text (dots). The full line is a fit of a double exponential $A_1 \exp(-t/\tau_1) + A_2 \exp(-t/\tau_2)$ to the data.
}
\label{fig:long_trace}
\end{figure}

Besides the relaxation mechanism discussed in the previous section, our data show evidence of another, much weaker relaxation process taking place on a longer time scale. In Fig.~\ref{fig:long_trace} we show an extended time trace after the heating pulse, averaged over one million repetitions (dots). The full line is obtained by fitting a double exponential of the form $A_1 \exp(-t/\tau_1) + A_2 \exp(-t/\tau_2)$ to the data. The fitted relaxation times are $\tau_1 = \SI{97}{\micro s}$ (the main relaxation) and $\tau_2 = \SI{0.41}{ms}$; the ratio between the two amplitudes is $A_2/A_1 = 0.018$. The origin of the slower relaxation process is presently unknown to us; however, the separation between the two time scales allows us to ignore the time dependence of the slower process during the thermal relaxation over $\tau_1$. For this reason, in the main text we fit a single exponential to the data with a corrected baseline. The baseline correction does not exceed $2\%$ in the data presented.

\section{Power optimization}

In order to measure the temperature sensitivity of our thermometer beyond linear-response, we proceed in the following manner. We first apply a continuous heating signal of varying amplitude $\IHpp$ to the island and measure temperature by using the thermometer in the liner response. Using the calibration of the resonator and the model for the NIS junction, we calibrate the island temperature against $\IHpp$. We then repeat the measurement for various input powers. Using the $\IHpp$-to-temperature calibration, we extract the temperature responsivity as $\partial P_{\rm det} / \partial T = (\partial P_\mathrm{det} / \partial \IHpp)(\partial \IHpp / \partial T)$, where $\Pdet$ is the measured mean power. After measuring the noise spectral density $S_{P_\mathrm{det}P_\mathrm{det}}$, we finally estimate the sensitivity as $\sqrt{S_{P_\mathrm{det}P_\mathrm{det}}}(\partial P_\mathrm{det} / \partial T)^{-1}$.

\begin{figure}
\centering
\includegraphics[width=12cm]{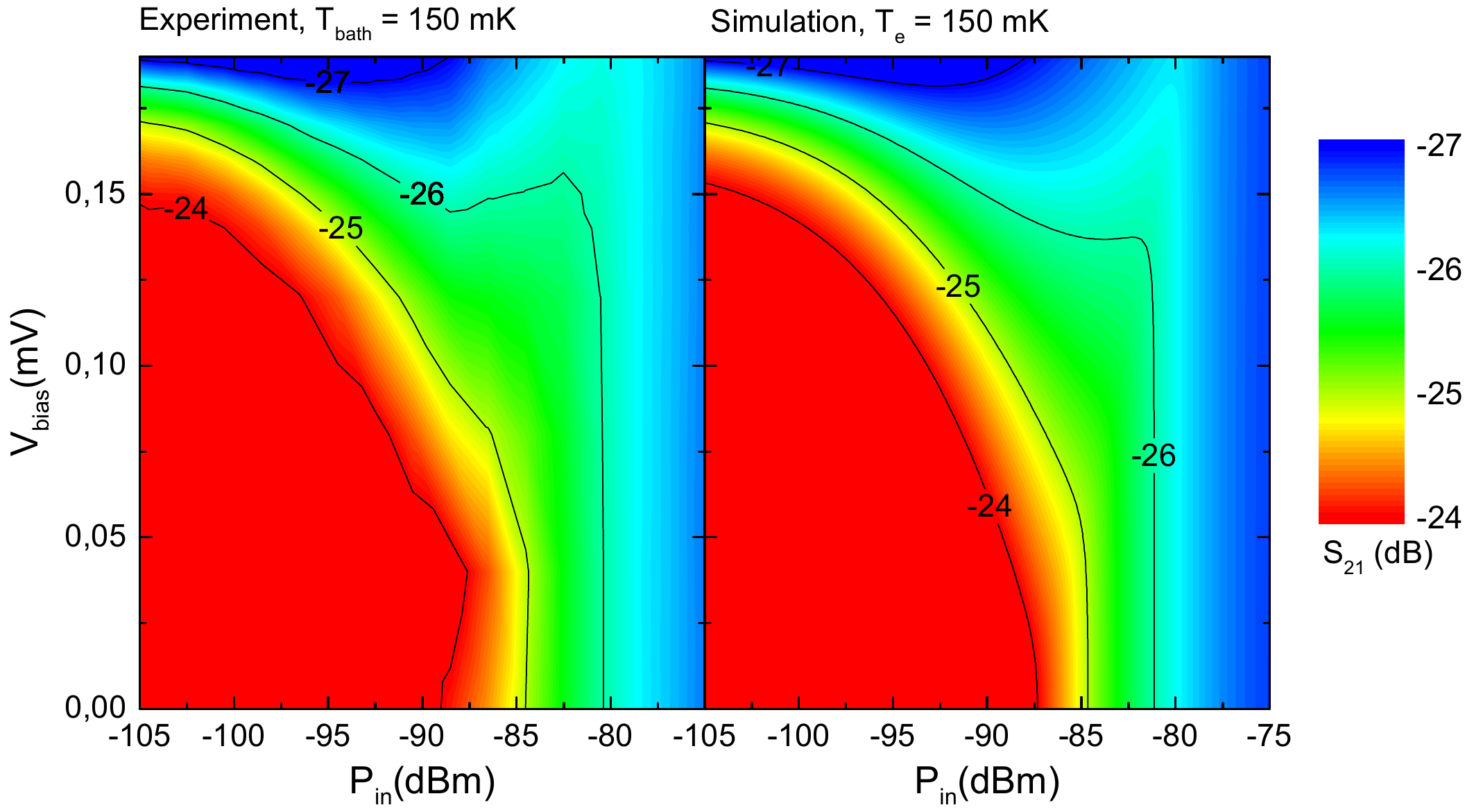}
\caption{Measured (left) and simulated (right) transmittance of power $|S_{21}|^2 = \Pdet/\Pgen$ as a function of $\Pin$ and $\Vb$. }
\label{fig:s21_map}
\end{figure}

In Fig.~4 in the main text, we compare the estimated sensitivity to numerical simulations. The simulations fully take into account the nonlinear current-voltage characteristics of the NIS junction and use the harmonic balance method to determine the response of the resonator terminated with the junction to a harmonic excitation of arbitrary amplitude. The power incident at the sample box $\Pin$ is obtained by subtracting the
total attenuation ($A$) of the input chain from the output power of the signal generator $\Pgen$.
Comparing the simulated $|s_{21}|^2$ with measurements as a function of $\Pin$ and $\Vb$ (see Fig.~\ref{fig:s21_map}) allows us to estimate $A~=~77.5\pm 1$~dB, and the gain of the output chain, $G = 55 \pm 1$~dB.

\section{Noise measurement}

We acquire real-time traces by demodulating the signal at the carrier frequency $f_0$ and recording the output with a fast digitizer. As a result, we obtain a power-versus-time trace over a bandwidth $B$ which is proportional to the sampling rate $f_S$.
If we assume the readout to be limited by the noise of our amplification chain, rather than by the intrinsic noise of our device, due to, e.~g., effective temperature fluctuations -- this assumption is verified {\it a posteriori} --, we can express $\Pdet$ and $S_{P_\mathrm{det}P_\mathrm{det}}$ as:
\begin{equation}
\begin{split}
\Pdet &= P_s + B G S_a \ , \\
S_{P_\mathrm{det}P_\mathrm{det}} &=  -2 B G^2 S_a^2 + 4  G S_a \Pdet \ , \label{eqs:noise}
\end{split}
\end{equation}
where $P_s$ is the signal without the noise and $S_a$ is the spectral density of the amplifier noise.
From \eqref{eqs:noise} we see that $\Pdet$ is offset by a constant amount, proportional to the bandwidth times the amplifier noise.
Furthermore, the noise $S_{PP}$ has a contribution which is proportional to $\Pdet$. 

\begin{figure}
\centering
\includegraphics{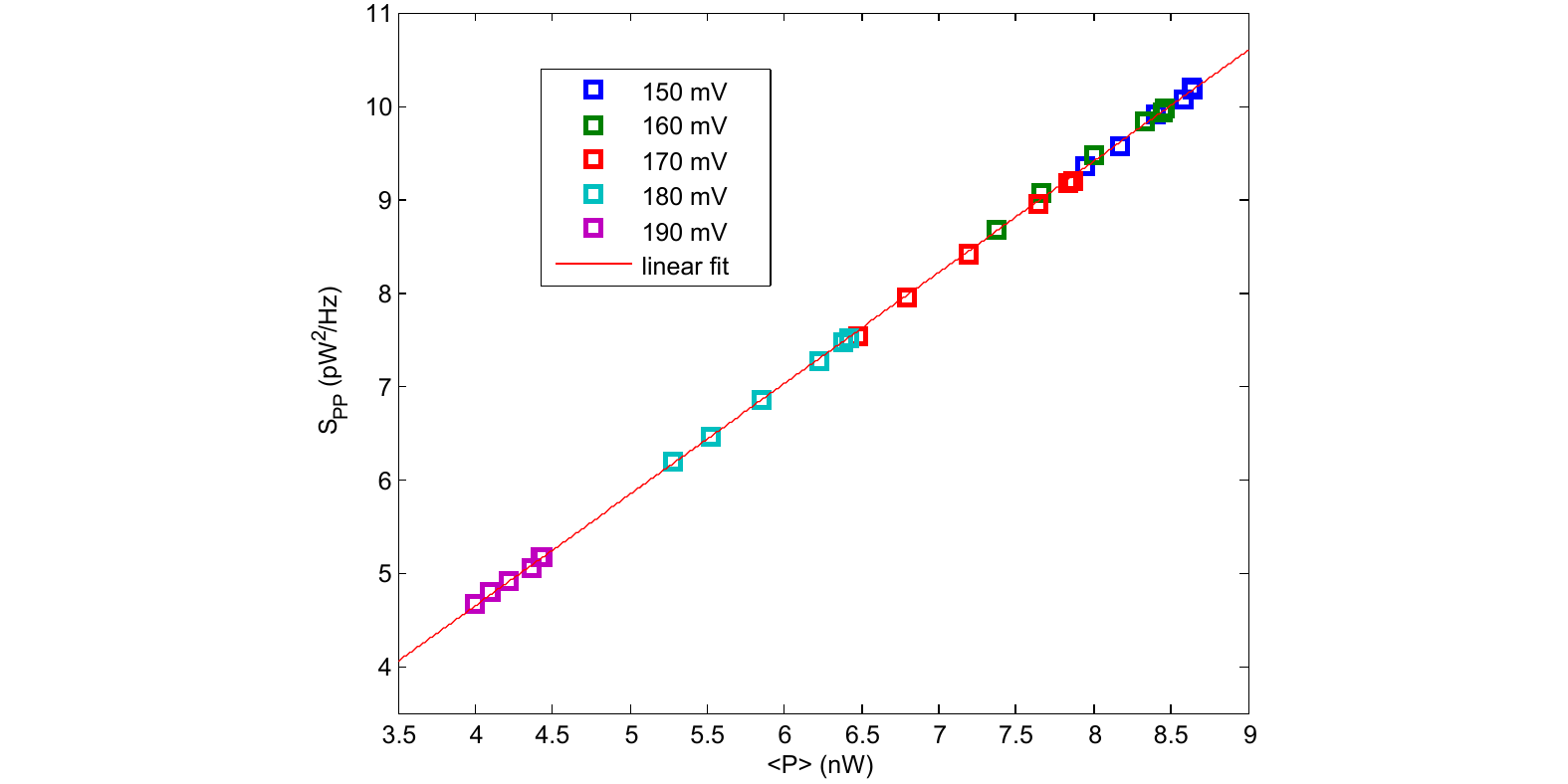}
%(a)
%\includegraphics[width=0.4\linewidth]{D:/Measurements/aaa_data_analysis/figs/fig_sampling_rate}
%\;
%(b)
%\includegraphics[width=0.4\linewidth]{D:/Measurements/aaa_data_analysis/figs/fig_noise_variance}
\caption{
%(a) Mean power $\Pdet$ versus sampling rate $f_S$. The measurement bandwidth $B$ is proportional to $f_S$.
%(b) 
Power noise spectral density $S_{P_\mathrm{det}P_\mathrm{det}}$ versus mean power $\Pdet$. The data are taken at $T_e=126$ mK for different bias voltages $\Vb$ (see the legend) and different input powers. The sampling rate is $f_s=\SI{0.5}{MS/s}$.}
\label{fig:noise}
\end{figure}

%In Fig.~\ref{fig:noise}(a) we measured $\Pdet$ for different values of the sampling rate. From a linear fit we extract
%$P_s = 1.54$ nW and $ \alpha G S_a =  4.6 \times 10^{-16}$ W/Hz, where $\alpha$ is the ratio between the noise equivalent bandwidth of the digitizer and the sampling rate (we expect $\alpha \gtrsim 1$).
In Fig.~\ref{fig:noise} we investigate the linear relationship between $S_{P_\mathrm{det}P_\mathrm{det}}$ and $\Pdet$ by measurements taken at different voltage biases and input powers. From a linear fit we extract $4 G S_a = 1.2\times 10^{-15}$ W/Hz, so that $G S_a = 3.0 \times 10^{-16}$ W/Hz.
The noise temperature of the chain is $T_\mathrm{sys}$ = $62 \pm 15$~K.
% despite the noise temperature of our amplifier being close to its nominal value of \SI{13.3}{K}.

\begin{thebibliography} {99}
\bibitem{schwab00} K. Schwab, E. A. Henriksen, J. M. Worlock, and M. L. Roukes, Nature  {\bf 404}, 974 (2000).
\bibitem{meschke06} M. Meschke, W. Guichard, and J. P. Pekola, Nature {\bf 444}, 187 (2006).
\bibitem{anthore13} S. Jezouin, F. D. Parmentier, A. Anthore, U. Gennser, A. Cavanna, Y. Jin, and F. Pierre, Science {\bf 342} 601 (2013).
\bibitem{ciliberto12}  A. B\'erut,	A. Arakelyan, A. Petrosyan, S. Ciliberto, R. Dillenschneider, and E. Lutz, Nature {\bf 483}, 187 (2012).	
\bibitem{toyabe10} S. Toyabe, T. Sagawa, M. Ueda, E. Muneyuki, and M. Sano, Nat. Phys. {\bf 6}, 988 (2010).
%\bibitem{koski14} J. V. Koski, V. F. Maisi, J. P. Pekola, and D. V. Averin, preprint available on arXiv:1402.5907.
\bibitem{koski14} J. V. Koski, V. F. Maisi, J. P. Pekola, and D. V. Averin, PNAS \textbf{111}, 13786 (2014).
\bibitem{pekola13} J. P. Pekola, P. Solinas, A. Shnirman, and D. V. Averin, New J. Phys. {\bf 15}, 115006 (2013).
%\bibitem{gasparinetti14} S. Gasparinetti, P. Solinas, A. Braggio, and M. Sassetti, arXiv:1404.3507.
\bibitem{gasparinetti14} S. Gasparinetti, P. Solinas, A. Braggio, and M. Sassetti, New J. Phys. \textbf{90}, 064505 (2014).
\bibitem{silaev14} M. Silaev, T. T. Heikkil\"a, and P. Virtanen, Phys. Rev. E {\bf 90}, 022103 (2014).
\bibitem{nahum95} M. Nahum and J. M. Martinis, Appl. Phys. Lett. {\bf 66}, 3203 (1995).
\bibitem{schmidt03} D. R. Schmidt, C. S. Yung, and A. N. Cleland, Appl. Phys. Lett. {\bf 83}, 1002 (2003).
\bibitem{schmidt04} D. R. Schmidt, C. S. Yung, and A. N. Cleland, Phys. Rev. B {\bf 69}, 140301(R) (2004).
\bibitem{schmidt05} D. R. Schmidt, K. W. Lehnert, A. M. Clark, W. D. Duncan, K. D. Irwin, N. Miller, and J. N. Ullom, Appl. Phys. Lett. {\bf 86}, 053505 (2005).
\bibitem{rowell76} J. M. Rowell and D. C. Tsui, Phys. Rev. B {\bf 14}, 2456 (1976).
\bibitem{nahum93}  M. Nahum and J. M. Martinis, Appl. Phys. Lett. {\bf 63}, 3075 (1993).
\bibitem{giazotto06} F. Giazotto, T. T. Heikkil\"a, A. Luukanen, A. M. Savin, and J. P. Pekola, Rev. Mod. Phys. {\bf 78}, 217 (2006).
\bibitem{schoelkopf98} R. J. Schoelkopf, P. Wahlgren, A. A. Kozhevnikov, and D. E. Prober, Science {\bf 280}, 1238 (1998).
\bibitem{qin06} H. Qin and D. A. Williams, Appl. Phys. Lett. {\bf 88}, 203506 (2006).
\bibitem{reilly07} D. J. Reilly, C. M. Marcus, M. P. Hanson, and A. C. Gossard, Appl. Phys. Lett. {\bf 91}, 162101 (2007).
\bibitem{saira12} O.-P. Saira, A. Kemppinen, V. F. Maisi, and J. P. Pekola, Phys. Rev. B {\bf 85}, 012504 (2012).
\bibitem{suppl} See Supplemental Material at [url].
\bibitem{pothier97} H. Pothier, S. Gu\'eron, N. O. Birge, D. Esteve, and M. H. Devoret, Phys. Rev. Lett. {\bf 79}, 3490 (1997).
\bibitem{muhonen12} J. Muhonen, M. Meschke, and J. P. Pekola, Rep. Prog. Phys. {\bf 75}, 046501 (2012).
\bibitem{nahum94} M. Nahum, T. M. Eiles, and J. M. Martinis, Appl. Phys. Lett. {\bf 65}, 3123 (1994).
\bibitem{rajauria09} S. Rajauria, H. Courtois, and B. Pannetier, Phys. Rev. B {\bf 80}, 214521 (2009).
\bibitem{roberts78} B.W. Roberts, {\it Properties of Selected Superconductive Materials}, NBS Technical Note 983, U.S Government Printing Office (1978).
\bibitem{peltonen10} J. T. Peltonen, P. Virtanen, M. Meschke, J. V. Koski, T. T. Heikkil\"a, and J. P. Pekola, Phys. Rev. Lett. {\bf 105}, 097004 (2010).
\bibitem{timofeev09} A. V. Timofeev, M. Helle, M. Meschke, M. M\"ott\"onen, and J. P. Pekola, Phys. Rev. Lett. {\bf 102}, 200801 (2009).
\bibitem{govenius14} J. Govenius, R. E. Lake, K. Y. Tan, V. Pietil\"a, J. K. Julin, I. J. Maasilta, P. Virtanen, and M. M\"ott\"onen, preprint available on arXiv:1403.6586 .
\bibitem{wellstood94} F. C. Wellstood, C. Urbina, and J. Clarke, Phys. Rev. B {\bf 49}, 5942 (1994).
\bibitem{taskinen04} L. J. Taskinen, J. M. Kivioja, J. T. Karvonen, and I. J. Maasilta, phys. stat. sol. (c) {\bf 1}, 2856 (2004).
\bibitem{karvonen05} J. T. Karvonen, L. J. Taskinen, and I. J. Maasilta, Phys. Rev. B {\bf 72}, 012302 (2005).
\bibitem{pobell} F. Pobell, {\it Matter and methods at low temperatures}, 3rd ed., Springer (2007).
\bibitem{anthore03} A. Anthore, F. Pierre, H. Pothier, and D. Esteve, Phys. Rev. Lett. {\bf 90}, 076806 (2003).
\bibitem{pascal13} L. M. A. Pascal, A. Fay, C. B. Winkelmann, and H. Courtois, Phys. Rev. B {\bf 88}, 100502 (2013).
\bibitem{long_tau_note} A small ($< 2\%$) baseline correction is applied to correct for a slow, much weaker relaxation process of unknown origin. See \cite{suppl} for details.
\bibitem{wallraff2004} A. Wallraff, D. I. Schuster, A. Blais, L. Frunzio, R. S. Huang, J. Majer, S. Kumar, S. M. Girvin, and R. J. Schoelkopf, Nature {\bf 431}, 162 (2004).
\bibitem{houck2007} A. A. Houck, D. I. Schuster, J. M. Gambetta, J. A. Schreier, B. R. Johnson, J. M. Chow, L. Frunzio, J. Majer, M. H. Devoret, S. M. Girvin, and R. J. Schoelkopf, Nature {\bf 449}, 328 (2007).
\bibitem{bozyigit2010} D. Bozyigit, C. Lang, L. Steffen, J. M. Fink, C. Eichler, M. Baur, R. Bianchetti, P. J. Leek, S. Filipp, M. P. da Silva, A. Blais, and A. Wallraff, Nat. Phys. {\bf 7}, 154 (2010).
\bibitem{karasik11} B. S. Karasik and R. Cantor, Appl. Phys. Lett. {\bf 98}, 193503 (2011).
\bibitem{karasik11a} B. S. Karasik, A. V. Sergeev, and D. E. Prober, IEEE Trans. Terahertz Sci. Technol. {\bf 1}, 97 (2011).
\bibitem{komiyama00} S. Komiyama, O. V. Astafiev, V. Antonov, T. Kutsuwa, and H. Hirai, Nature 403, 405 (2000).
\bibitem{komiyama11} S. Komiyama, IEEE J. Sel. Top. Quantum Electron. 17, 54 (2011).
\bibitem{castellanos-beltran07} M. A. Castellanos-Beltran and K. W. Lehnert, Appl. Phys. Lett. {\bf 91}, 083509 (2007).
\end{thebibliography}
\end{document}